\documentclass[a4paper,11pt]{article}
\usepackage{pos}

\usepackage{graphicx}
\usepackage{braket}
\usepackage{multirow}
\usepackage{url}
\usepackage{hvfloat}
\usepackage{commath}
\usepackage{amsmath}
\usepackage{caption}
\usepackage{booktabs}
\usepackage{colortbl}
\usepackage{comment}
\usepackage[export]{adjustbox}
\usepackage{sidecap}
\usepackage{wrapfig}
\usepackage{paralist}
\usepackage{enumitem}
\usepackage[nolist,nohyperlinks]{acronym}
\usepackage{blindtext}
\usepackage{lipsum}
\usepackage[left]{lineno}
\usepackage[symbol,bottom]{footmisc}
\renewcommand{\thefootnote}{\fnsymbol{footnote}}

\def\figureautorefname~#1\null{Fig.\,#1\null}
\usepackage{titlesec}
\titlespacing*{\section}
  {0pt}{2.0ex plus 0.5ex minus 0.2ex}{0.8ex plus 0.3ex}
\titlespacing*{\subsection}
  {0pt}{0.8ex plus 0.4ex minus 0.1ex}{0.6ex plus 0.2ex}
  
\DeclareMathSizes{10}{9}{6}{5}

\newlength{\bibitemsep}
\newlength{\bibparskip}
\let\oldthebibliography\thebibliography
\renewcommand\thebibliography[1]{%
  \oldthebibliography{#1}%
  \setlength{\parskip}{\bibitemsep}%
  \setlength{\itemsep}{\bibparskip}%
}
\setlist[enumerate]{%
  leftmargin=1.5em,   
  labelsep=0.5em      
}

\acrodef{UHE}{ultra-high-energy}
\acrodefplural{UHE}{ultra-high-energies}
\acrodef{UHECR}{ultra-high-energy cosmic ray}
\acrodefplural{UHECR}{ultra-high-energy cosmic rays}
\acrodef{GMF}{Galactic Magnetic Field}
\acrodef{EGMF}{Extragalactic Magnetic Field}
\acrodef{EB}{Editorial Board}
\acrodefplural{EB}{Editorial Boards}
\acrodef{EAS}{extensive air shower}
\acrodefplural{EAS}{extensive air showers}
\acrodef{SGP}{supergalactic plane}
\acrodef{FD}{Fluorescence Detector}
\acrodef{SD}{Surface Detector}
\acrodefplural{SD}{Surface Detectors}
\acrodef{SBG}{starburst galaxies}
\acrodef{AGN}{active galactic nuclei}
\acrodef{DNN}{Deep Neural Network} 
\acrodefplural{DNN}{Deep Neural Networks}
\acrodef{1DCNN}[1-D CNN]{one dimensional Convolutional Neural Network} 
\acrodefplural{1DCNN}[1-D CNNs]{one dimensional Convolutional Neural Networks} 
\acrodef{CNN}[CNN]{Convolutional Neural Network} 
\acrodefplural{CNN}[CNNs]{Convolutional Neural Networks} 
\acrodef{LSTM}{Long Short-Term Memory}
\acrodefplural{LSTM}[LSTMs]{Long Short-Term Memory networks}
\acrodef{HPC}{High Performance Computing Center}
\acrodef{MURF}{Mines Undergraduate Research Fellowship}
\acrodef{ML}{Machine Learning}
\acrodef{HIM}{Hadronic Interaction Model}
\acrodefplural{HIM}[HIMs]{Hadronic Interaction Models}
\acrodef{GFN}[GFN]{Great Flexible Network}
\acrodef{MSE}[MSE]{Mean Squared Error}
\acrodef{WCD}[WCD]{Water Cherenkov Detector}
\acrodefplural{WCD}[WCDs]{Water Cherenkov Detectors}
\acrodef{SSD}[SSD]{Surface Scintillator Detector}
\acrodefplural{SSD}[SSDs]{Surface Scintillator Detectors}
\acrodef{AERA}[AERA]{Auger Engineering Radio Array}
\acrodef{RD}[RD]{Radio Detector}
\acrodefplural{RD}[RDs]{Radio Detectors}
\acrodef{HEAT}[HEAT]{High Elevation Auger Telescopes}
\acrodef{TS}[TS]{Test Statistic}
\acrodef{POEMMA}[POEMMA]{Probe of Multi-Messenger Astrophysics}
\acrodef{PBR}[PBR]{POEMMA Balloon with Radio}
\acrodef{VHENu}[VHE-$\nu$]{Very-High-Energy Neutrinos}
\acrodef{HAHA}[HAHA]{High-Altitude Horizontal Air-Shower}
\acrodefplural{HAHA}[HAHAs]{High-Altitude Horizontal Air-Showers}
\acrodef{ToO}[ToO]{Target of Opportunity}
\acrodefplural{ToO}[ToOs]{Targets of Opportunity}
\acrodef{SPB}[SPB]{Super Pressure Balloon}
\acrodefplural{SPB}[SPBs]{Super Pressure Balloons}
\acrodef{PUEO}[PUEO]{the Payload for Ultrahigh Energy Observations}
\acrodef{SIP}[SIP]{Support Instrumentation Package}
\acrodef{CC}[CC]{Cherenkov Camera}
\acrodef{FC}[FC]{Fluorescence Camera}
\acrodef{RoC}[RoC]{Radius of Curvature}
\acrodef{ACP}[ACP]{Aspheric Corrector Plate}
\acrodef{FoV}[FoV]{Field-of-View}
\acrodef{PDM}[PDM]{Photo Detection Module}
\acrodefplural{PDM}[PDMs]{Photo Detection Modules}
\acrodef{GFRV}[GFRV]{Glass Fiber Reinforced Vinylester}
\acrodef{FR4}[FR4]{Flame Retardant 4 glass-reinforced epoxy laminate}
\acrodef{G10}[G10]{Glass-reinforced epoxy laminate \#10}
\acrodef{EMI}[EMI]{Electromagnetic Interference}
\acrodef{CSBF}[CSBF]{Columbia Scientific Balloon Facility}

\title{The Optical and Mechanical Design of POEMMA Balloon with Radio}
\ShortTitle{PBR Optomechanics}

\author*[a]{\small Eric Mayotte}
\author[b]{\small Austin Cummings}
\author[c]{\small Paul Degarate}
\author[d]{\small Neville DeWitt Pierrat}
\author[e]{\small Johannes Eser}
\author[f]{\small William Finch}
\author[a]{\small Julia Burton-Heibges}
\author[a]{\small Tobias Heibges}
\author[g]{\small Eric Mentzell}
\author[h]{\small Stephan Meyer}
\author[a]{\small Conrad Shay}
\author[h]{\small Benjamin Stillwell}
\author[i]{\small Yoshiyuki Takizawa}
\author[a]{\small Luke Wanner}
\author[a]{\small Lawrence Wiencke}
\onbehalf{\small for the JEM-EUSO Collaboration}

\affiliation[a]{Department of Physics, Colorado School of Mines, 1523 Illinois St., Golden, CO, USA}
\affiliation[b]{Department of Physics, Pennsylvania State University, 251 Pollock Rd, University Park, PA, USA}
\affiliation[c]{Halley Tech Inc, 2257 Arlington Ct, Auburn Al, USA}
\affiliation[d]{Firebrand Engineering, \texttt{contact@firebrand.engineering}, Golden CO, USA}
\affiliation[e]{Columbia Astrophysics Laboratory, Columbia University, 538 West 120th Street, New York, NY, USA}
\affiliation[f]{Cotopaxi Mech, \texttt{https://cotopaximech.com}, Denver CO, USA}
\affiliation[g]{NASA Goddard Space Flight Center, 8800 Greenbelt Rd, Greenbelt, MD, USA}
\affiliation[h]{Department of Astronomy \& Astrophysics, University of Chicago, 5640 S Ellis Ave, Chicago, IL, USA}
\affiliation[i]{Center for Advanced Photonics, RIKEN, 2-1 Hirosawa, Wako, Saitama, Japan}

\emailAdd{emayotte@mines.edu}

\abstract{POEMMA Balloon with Radio (PBR) is a NASA super-pressure balloon mission building toward the proposed Probe Of Extreme Multi-Messenger Astrophysics (POEMMA) dual satellite mission. 
In its planned 2027 launch, PBR will study Ultra-High-Energy Cosmic Rays, Neutrinos, and High-Altitude Horizontal Airshowers from 33\,km above the Earth. 
By operating at balloon altitudes, PBR will provide a novel vantage point to study air-shower physics while offering competitive instantaneous exposure to neutrinos from transient astrophysical phenomena. 
The payload's optical instrument is a 0.95\,m\textsuperscript{2} aperture hybrid Schmidt telescope with a 3.81\,m\textsuperscript{2} segmented mirror focusing light onto a Fluorescence Camera and a bi-focalized Cherenkov Camera. 
The payload will also feature a Radio Instrument consisting of two sinuous antennas based on the Payload for Ultrahigh Energy Observations (PUEO) low-frequency instrument. 
A combined gamma ray/x-ray detector and IR cloud camera round out the instrumentation package, meaning PBR will be the first multi-hybrid balloon-borne multi-messenger observatory flown. 
This extensive instrumentation must be combined into a radio quiet payload that satisfies the scientific needs and can operate in near vacuum at extreme temperatures, all while meeting NASA safety requirements and weighing no more than 3000\,lbs (1361\,kg). 
Accomplishing these tasks together will mark a significant step toward establishing technological readiness for the POEMMA satellite mission. 
We present an overview of PBR’s mechanical and optical systems, additionally detailing our strategies to mitigate electromagnetic interference for the radio instrument and prepare for the harsh near-space environment.}

\ConferenceLogo{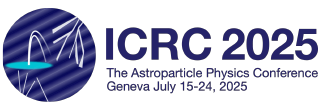}

\FullConference{39th International Cosmic Ray Conference (ICRC2025)\\
 15–24 July 2025\\
Geneva, Switzerland\\}

\begin{document}
\maketitle

\section{POEMMA Balloon with Radio}\label{sec:intro}

The study of \acp{UHECR} and \ac{VHENu} has progressed significantly over the last twenty years, driven by the high precision and large exposures of modern ground-based observatories. 
However, fundamental questions about their origins and nature remain unanswered, particularly at the highest energies. 
Resolving these will likely require next-generation observatories offering substantially improved resolution and/or orders-of-magnitude larger exposure~\cite{Coleman:2022abf}. 
Space-based instruments present a practical route to such exposure, surveying vast atmospheric volumes from a single platform. 
The proposed dual-satellite \ac{POEMMA} mission aims to extend the exposure limits for both \acp{UHECR} and \acp{VHENu} simultaneously by exploiting this approach~\cite{POEMMA:2020ykm}.

In the development of space-based observatories, pathfinder balloon missions are crucial for validating detection techniques and advancing technological readiness prior to full-scale satellite deployments. 
\ac{PBR} is the third and most advanced \ac{SPB} pathfinder mission undertaken by the JEM-EUSO collaboration and is planned to fly from New Zealand in early 2027 (for a detailed overview, see~\cite{Johannes:ICRC}). 
Leveraging the successes and experience from previous balloon missions, EUSO-SPB1~\cite{JEM-EUSO:2023ypf} and EUSO-SPB2~\cite{Adams:2025owi}, \ac{PBR} significantly upgrades observational capabilities compared to earlier balloon payloads. 
\ac{PBR} incorporates a hybrid focal surface combining a \ac{FC}~\cite{Francesco:ICRC} and \ac{CC}~\cite{Valentina:ICRC} into a Schmidt telescope, following the design proposed for \ac{POEMMA}.
\ac{PBR} also uniquely adds simultaneous radio, X-ray, and $\gamma$-ray detection capabilities allowing for multi-channel multi-hybrid observations of air showers~\cite{Matteo:ICRC}. 
Furthermore, the \ac{PBR} telescope can rotate freely in azimuth and point from nadir to $+15^\circ$ above the horizon to target different phenomena and follow up on astrophysical alerts. 
This combination of instrumentation makes \ac{PBR} the first multi-hybrid high-energy stratospheric astroparticle observatory, enabling it to pursue its primary science goals:
\begin{enumerate}[topsep=3pt]\setlength\itemsep{-0.1em}
    \item \textbf{Investigating the Origin of UHECRs:} 
        \ac{PBR} will perform the first sub-orbital measurement of \ac{EAS} fluorescence induced by \acp{UHECR}, validating key detection technologies planned for \ac{POEMMA}.
    \item \textbf{Studying \acp{HAHA}:}
        \ac{PBR}'s unique stratospheric vantage point will enable the first extensive study of rarely observed \acp{HAHA}. 
        Simultaneous detection through optical Cherenkov, radio, X-ray, and $\gamma$-ray instruments will provide a multi-hybrid characterization of highly extended air showers developing in rarefied atmospheres, facilitating precise measurements of cosmic-ray spectrum near PeV energies.
    \item \textbf{Searching for Astrophysical Neutrinos:}
        \ac{PBR} will provide high instantaneous exposure to Earth-skimming tau neutrinos via below-horizon \ac{EAS} detections using Cherenkov and radio instruments. 
        This capability enables fast follow-ups of transient multi-messenger sources.
\end{enumerate}
\section{The General Design of the PBR Payload}
\label{sec:payload}

The \ac{PBR} payload comprises the mechanical structure and scientific instrumentation package carried aboard the NASA \ac{SPB}. 
It hosts all instruments essential to achieving \ac{PBR}'s scientific goals, while conforming to rigorous NASA flight and safety requirements. 
Each component must withstand conditions such as pressures down to 1\,mbar, temperatures as low as $-70^\circ$\,C, and accelerations exceeding 8\,G with considerable margins of safety. 
Although the standard maximum payload weight for NASA balloons is 2500\,lbs (1134\,kg), \ac{PBR} has been granted an exception, increasing its allowable science payload to 3000\,lbs (1361\,kg).
Additionally, minimizing electromagnetic interference is critical for the successful operation of the sensitive radio instrument.
Meeting these constraints simultaneously presents significant engineering challenges.

\begin{wrapfigure}{r}{0.55\linewidth}
  \vspace*{-3mm} 
    \centering
    \includegraphics[width=\linewidth]{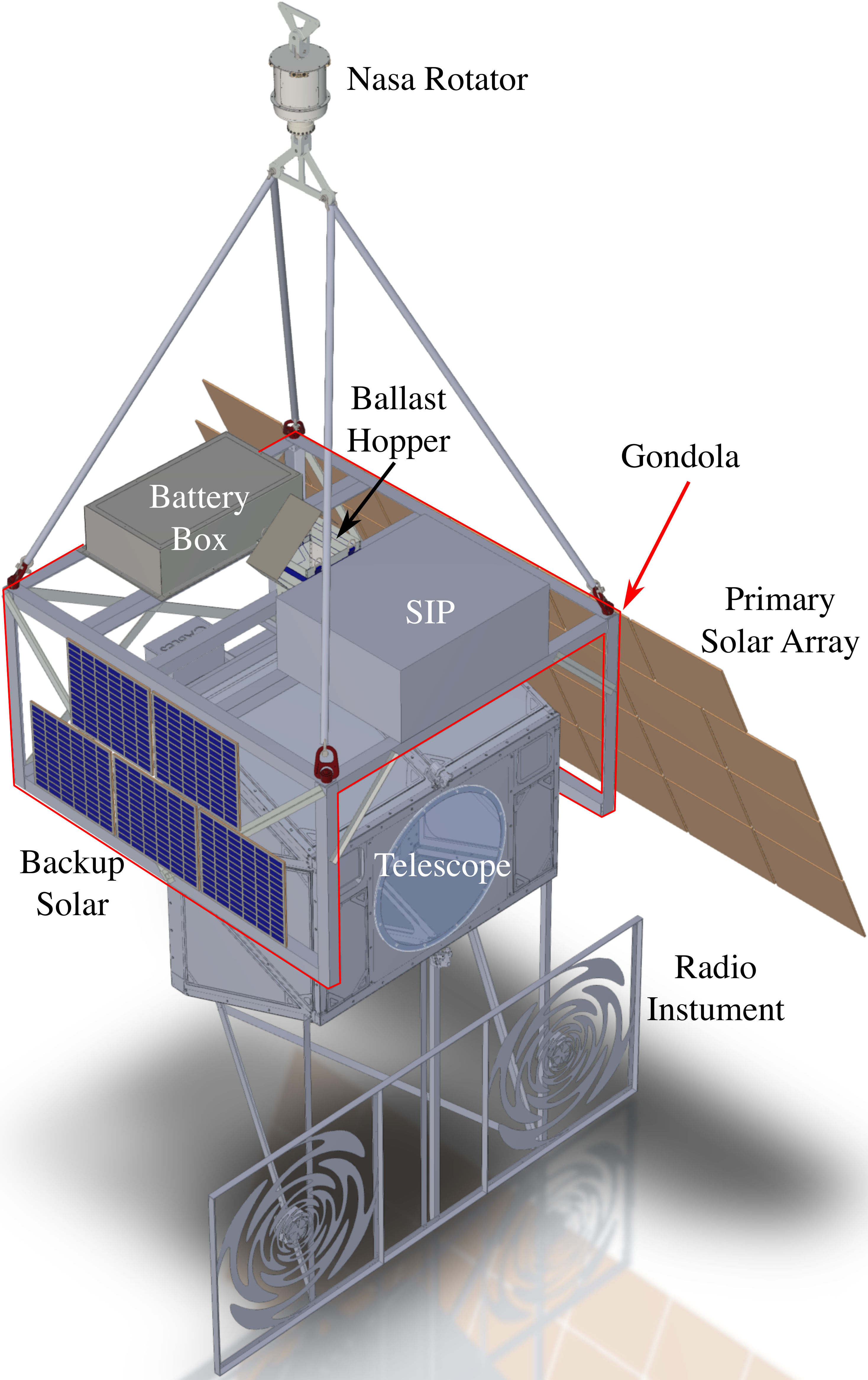}
    \caption{Overview of the PBR Payload}
    \label{fig:PayloadOverview}
  \vspace*{-3mm} 
\end{wrapfigure}
\autoref{fig:PayloadOverview} provides a schematic overview of the payload's layout and main components. 
The \textbf{Rotator} connects the super-pressure balloon to the payload gondola, allowing precise azimuthal rotation control to $1^\circ$ accuracy~\cite{Burth2022_5p5KRotator}, with absolute pointing determined with sub-$1^\circ$ accuracy via differential GPS. 
The \textbf{Gondola} provides the structural framework and mounting points for all payload subsystems. 
It is made out of aerospace-grade (7075) Aluminum alloy angle with steel nut blocks and rigging points for a high-strength, low-weight combination assembly. 
The \textbf{Power System} includes a \emph{Battery Box}, housing seven high-capacity batteries, heaters, and power-management electronics; a \emph{Primary Solar Array}, consisting of 27 lightweight, high-efficiency panels that actively track the sun via the Rotator; and a \emph{Backup Solar Array} of five fixed panels to ensure safe operation in case of Rotator failure.

The \textbf{Ballast Hopper}, part of NASA's standard systems, stores and dispenses up to 600\,lbs (272\,kg) of ballast to regulate balloon altitude. 
NASA also provides the \textbf{\ac{SIP}}, which handles telemetry, communication, and command functions. 
The \textbf{Telescope} is the primary scientific instrument (see \autoref{sec:Telescope}), housing the optical system (\autoref{sec:optics}), \ac{CC}\cite{Valentina:ICRC}, \ac{FC}\cite{Francesco:ICRC}, X-ray/$\gamma$-ray detector~\cite{Matteo:ICRC}, an IR cloud camera, calibration equipment, and instrument control computers~\cite{Valentina2:ICRC}.

The \textbf{Radio Instrument} includes two antennas modeled after \ac{PUEO}'s low-frequency antenna~\cite{Abarr_2021}, along with their support structures and deployment mast (see \autoref{sec:RI}). 
This instrument operates in conjunction with the Cherenkov detector, measuring radio emissions from \acp{EAS} in the frequency range from approximately 50 to 500\,MHz. 
Lastly, the \textbf{Vertical Rotator} adjusts the elevation pointing of the Telescope and Radio Instrument from nadir to $+15^\circ$ above the horizon, achieving steering accuracy around $0.1^\circ$, with precise pointing direction verified by onboard inclinometers~\cite{Lawrence:ICRC}.
\section{PBR Optics}\label{sec:optics}

\begin{figure}[!b]
    \centering
    \begin{minipage}[t]{0.48\linewidth}
        \centering
        \includegraphics[width=\linewidth]{figures/OpticalDiagram.pdf}
        \caption{The Optical Sketch of PBR}
        \label{fig:OpticalPrescription}
    \end{minipage}%
    \hfill
    \begin{minipage}[t]{0.48\linewidth}
        \centering
        \includegraphics[width=\linewidth]{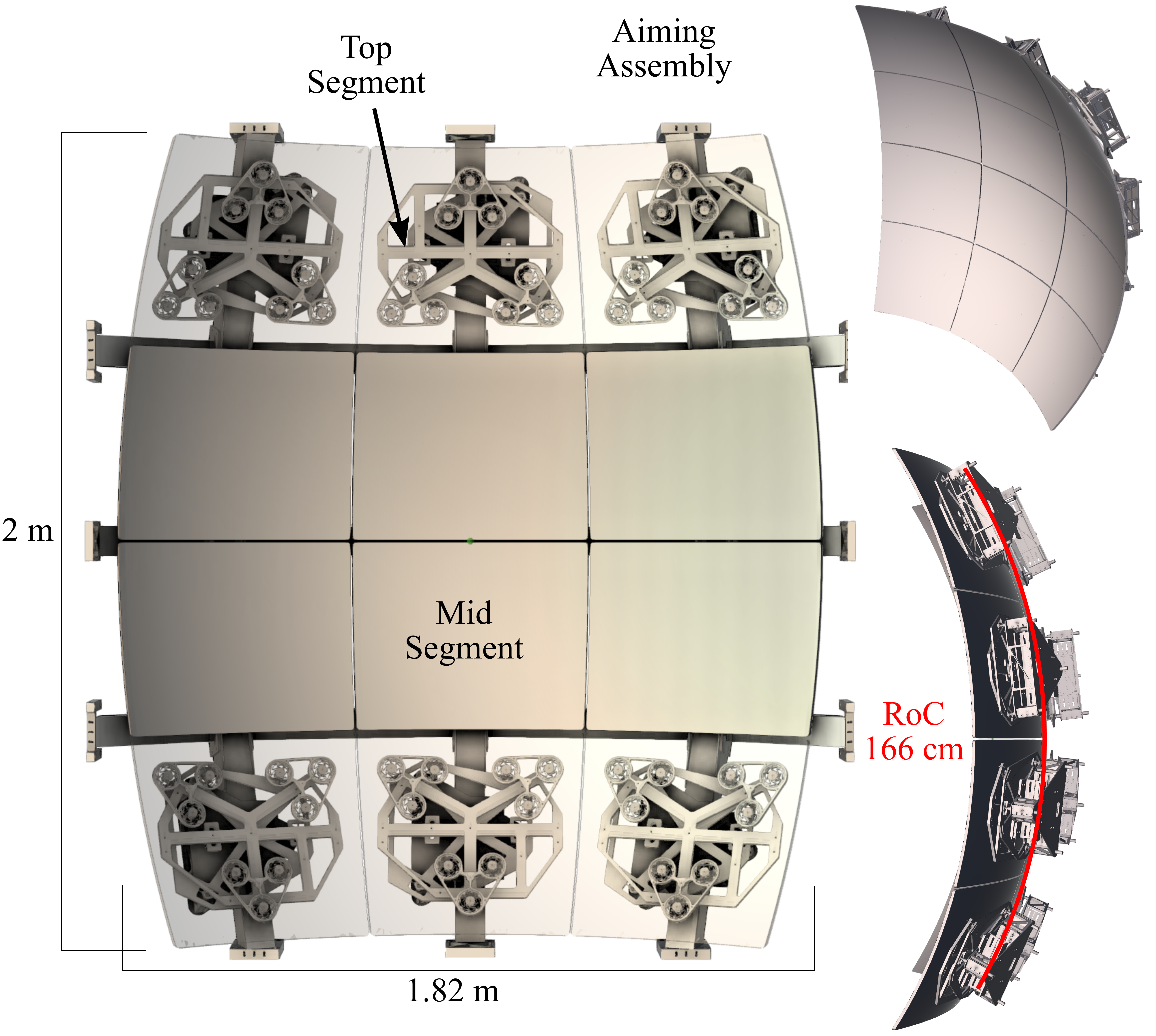}
        \caption{The Primary Mirror of PBR}
        \label{fig:Mirror}
    \end{minipage}
\end{figure}

The \ac{PBR} telescope is of Schmidt design for several reasons.
Schmidt telescopes offer a fast focal ratio, allowing the telescope to accommodate a large aperture within a compact and lightweight assembly.
They provide a wide, well-corrected \ac{FoV} with few optical parts, maximizing sensitivity and ensuring uniform image quality across \ac{PBR}'s large hybrid focal surface. 
Schmidt telescopes also allow for the use of a spherically curved mirror, significantly simplifying the mirror design and fabrication costs, while also simplifying the mechanical design of the mirror support frame.
Additionally, only a thin \ac{ACP} is required, which keeps the payload mass and volume low. 
Last, their straightforward and squat design allows the sturdy construction needed to minimize flexing without adding unmanageable weight to the payload.
An optical diagram of the \ac{PBR} Schmidt telescope is shown in \autoref{fig:OpticalPrescription}.

The \ac{PBR} telescope needs to support a $27^\circ \times 36^\circ$ \ac{FoV} and a 1.1\,m diameter, 0.95\,m$^2$ aperture.
As shown in \autoref{fig:Mirror}, to achieve this, it features an approximately $1.82$\,m $\times$ $2$\,m, $3.81$\,m$^2$ segmented spherical mirror with a $\sim1.66$\,m \ac{RoC} and a large \ac{ACP} to minimize distortion at the edges of the \ac{FoV}.
As shown in \autoref{fig:Mirror}, the mirror is composed of 12 trapezoidal segments with a slightly different geometry for the middle rows as compared to the top and bottom rows.
All segments are made of 11\,mm-thick, vacuum-slumped, borosilicate glass with a vacuum-deposited aluminum film layer optimized for reflectivity in the UV, which is protected by a silicon dioxide coating. 
Both mirror segments weigh $\sim 18$\,lbs (8.16\,kg).
The other main optical component is the polymethyl methacrylate (PMMA) \ac{ACP}, which sits at the aperture to correct distortions at the edge of \ac{FoV}.
It has been shaped in a precision diamond-turning process to a thickness of 10\,mm in the center, which tapers to $\sim 5$\,mm at its lowest point near the edges, after which it is polished.

To fit the \ac{CC} and \ac{FC}, the $27^\circ \times 36^\circ$ \ac{FoV} focal surface is approximately $38$\,cm $\times$ $51.4$\,cm in size and has an \ac{RoC} of $\sim 853$\,mm.
The four \acp{PDM} of the \ac{FC} occupy the top focal surface with a $27^\circ \times 27^\circ$ \ac{FoV}.
Because the detection surface of each \ac{FC} \ac{PDM} is flat, a field flattener is placed in front of each to map the spherical focal surface to the \ac{PDM} face.
To limit background contamination, each \ac{PDM} also has a BG3 filter placed over the detection surface, allowing only wavelengths between 290\,nm and 430\,nm to reach the camera.

\renewcommand\sidecaptionsep{2.2mm}
\sidecaptionvpos{figure}{c}
\begin{SCfigure}[][!t]
    \vspace{-1mm}
    \centering
    \includegraphics[width = 0.7\textwidth]{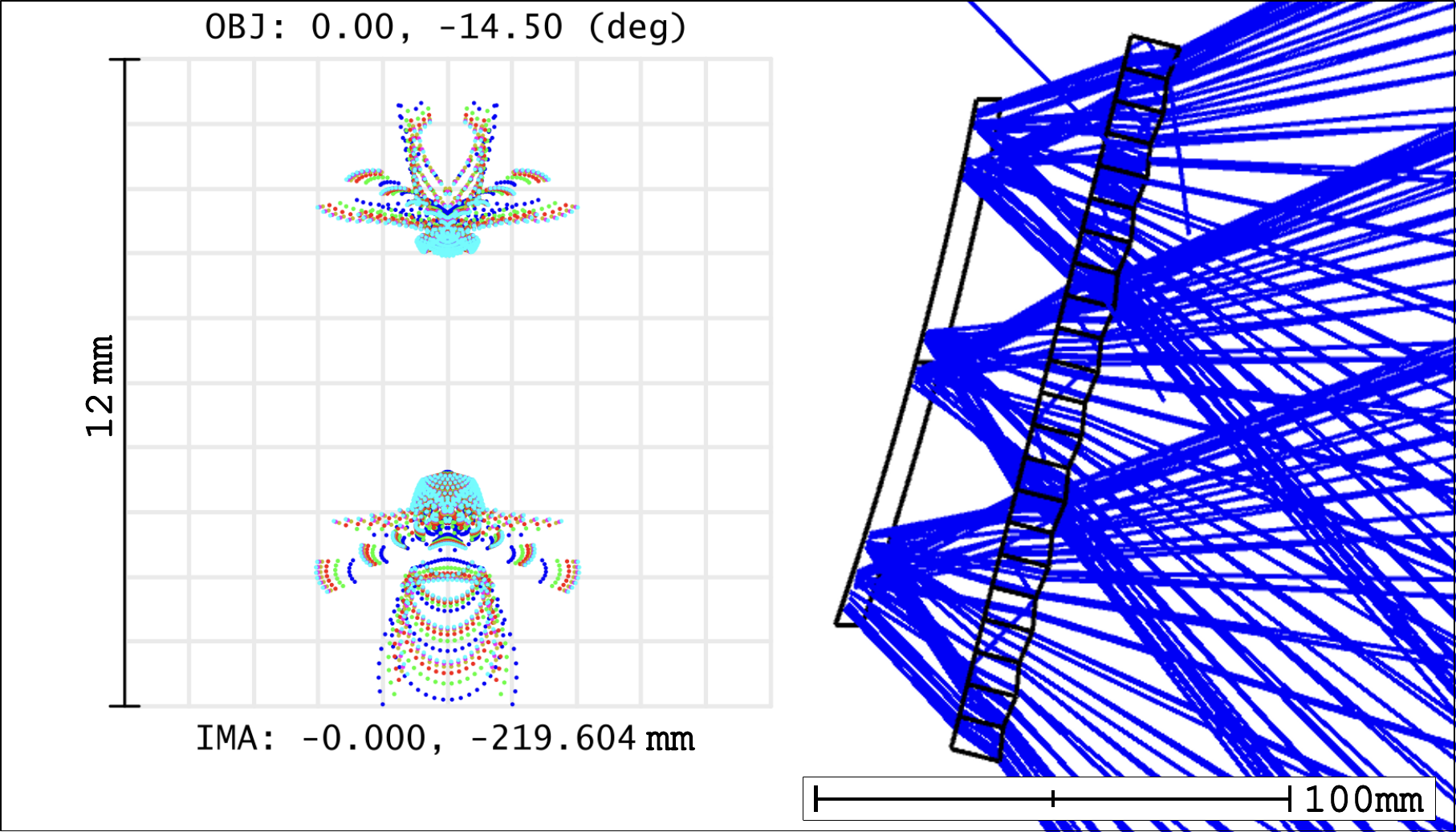}
    \caption{The Cherenkov Camera Bifocalization Optics. Right: Zemax simulation of a prototype diagram of a vertical biafocalization optic placed in front of the focal surface of the \ac{CC}. Left: spot diagram from a bifocalizer built to produce a $\sim6$\,mm spot separation (2 pixels) showing the majority of light is well contained within a single $3\times3$\,mm pixel.}
    \label{fig:BiFocalizer}
\end{SCfigure}

The \ac{CC} sits at the bottom of the focal surface, occupying $12^\circ \times 6^\circ$ of the \ac{FoV}. 
The \ac{CC} features a curved detection face which closely follows the focal surface \ac{RoC}.
As shown in \autoref{fig:BiFocalizer}, in the current design, bifocalization optics (Bifocalizer) are positioned in front of the \ac{CC} to split the optical signal into two well-separated pixels.
The performance of the Bifocalizer is being evaluated.
In case the performance is insufficient for reliable low-threshold triggering in the \ac{CC}, aiming the mirror segments to create the two desired spots on the \ac{CC} is also being explored.
The optical signal of the \ac{CC} is split into two non-adjacent pixels to provide a clear veto of direct charge particle strikes to the camera, as only light that reaches the telescope nearly parallel will be split, indicating that the event occurred far from the telescope.
The remaining angular space of the focal surface is taken up by the mechanical requirements of holding and supporting the cameras and optics.
\section{PBR Telescope Mechanics}\label{sec:Telescope}

\begin{figure}[!htb]
    \centering
    \includegraphics[width=.8\linewidth]{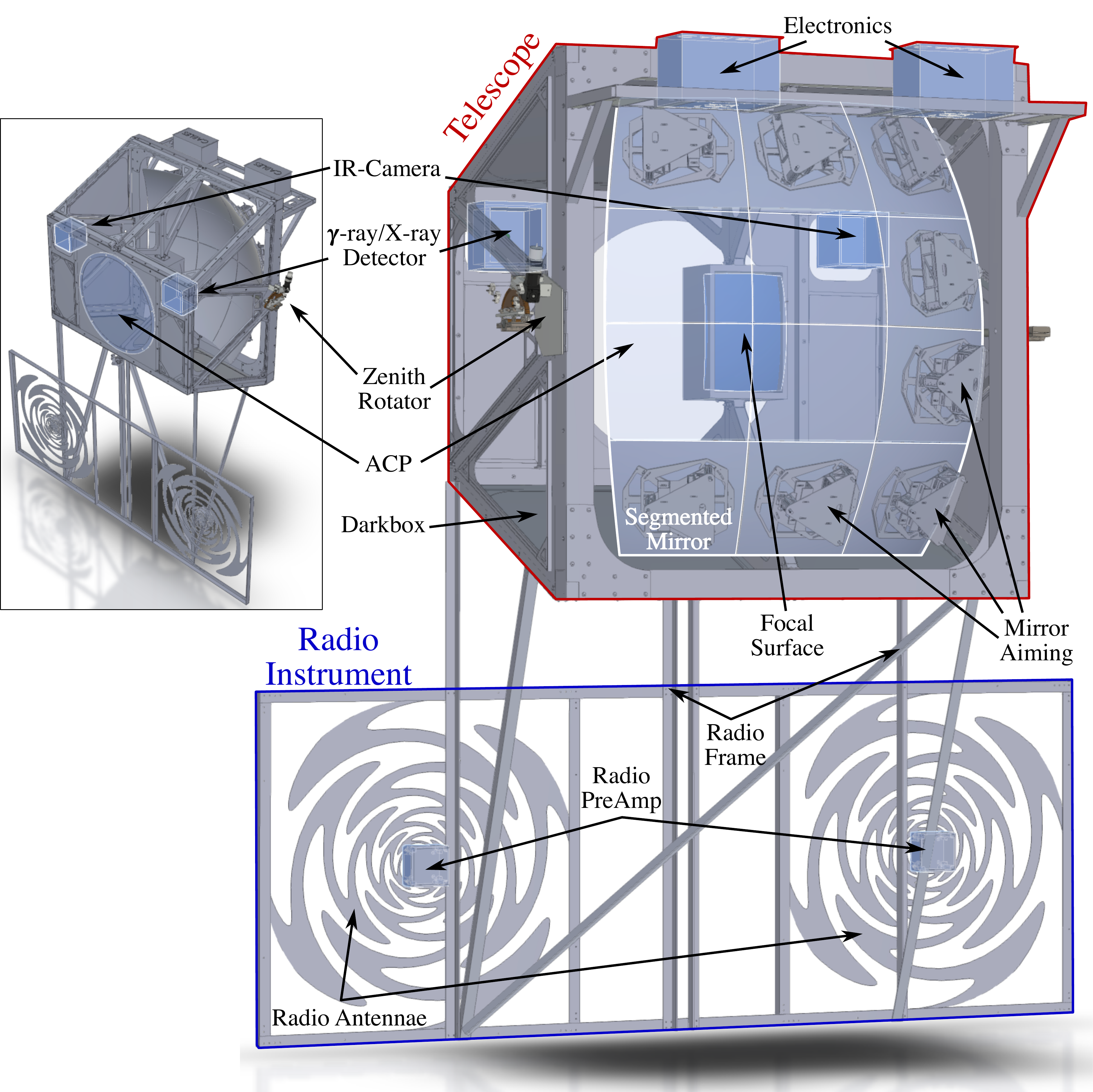}
    \caption{The Telescope Assembly of PBR}
    \label{fig:Telescope}
\end{figure}

The mechanical structure of the telescope was designed to support the optical prescription described above.
Like any steerable telescope, the structure had to be rigid enough to prevent the sagging of any optical components due to gravity when moving over its full range of pointing.
In contrast to any telescope placed on the ground, except for in the most extreme circumstances, the telescope of \ac{PBR} must also preserve optical performance and structural integrity over a temperature range of $-70^\circ$\,C to $25^\circ$\,C and accelerations has high as 8\,G with considerable safety factors.
Furthermore, it must survive in near-vacuum and high-UV environments without the ability to repair any components that fail during flight.
The structure designed to meet these challenges is shown in \autoref{fig:Telescope}, with the rear mirror support cage (shown in \autoref{fig:Mirror}) removed for visibility.

The ruling mechanical constraint is the optical stiffness of the assembly, as this is a higher bar than the yield constraints imposed by NASA.
Since 6061 and 7075 Aluminum alloys have very similar Young's moduli, and 6061 is cheaper, easier to procure, and is available in more form factors, 6061 has been chosen as the alloy used for the majority of the telescope.
The exception applies to the flexure assemblies described below and a select few other components, where yield was observed within the safety margins on the upper end of the possible acceleration range.
The telescope structure can be broken down into five major components: the mirror assembly, the telescope frame, the aperture assembly, the camera shelf, and the dark-box shell.

\paragraph*{Telescope Frame}
The core of the telescope frame is the large square belt frame, which wraps vertically around the mirror, just behind the center of mass of the telescope.
The belt frame bears the majority of the telescope's load, stiffens it, and directly interfaces with the rotator system. 
The Mirror Assembly bolts directly to this belt, providing very high stiffness to minimize sag in any aiming direction.
From this belt, two trusses extend forward on both sides of the belt frame to hold the aperture assembly in place. 
In the center of the telescope, an additional smaller truss extends forward to the aperture assembly to hold the camera shelf in place. 
On the outside of these trusses, gussets are built in to stiffen the assembly and hold the panels of the dark box shell.
Due to weight concerns, the material of the telescope frame has been minimized in all recognized opportunities. 

\paragraph*{Aperture Assembly}
The Aperture Assembly must do three main things: it must hold the \ac{ACP} firmly in place, it must hold and orient the IR-Camera and $\gamma$/X-ray detector (indicated in \autoref{fig:Telescope}, and it must hold the shutters (seen in \autoref{fig:OpticalPrescription}.
To maintain the required stiffness, the front of the aperture assembly is skinned with an Aluminum sheet with peepholes for the IR-Camera and $\gamma$/X-ray detector.
The shutters are mounted to the top and bottom of the assembly on the central trusses. 

\paragraph*{Dark-Box shell}
The Dark box panels are composed of a layered structure.
Reflective Mylar forms the outermost layer, reflecting solar radiation and keeping the telescope temperatures low during the day.
Under the Mylar sits a layer of Titan RF Faraday fabric~\cite{titanrf_fabric} to transform the telescope into a Faraday cage, thereby suppressing EMI from the various electronics within the telescope. 
Next is a 19\,mm thick layer of high-density extruded polyurethane foam to insulate the telescope from extreme cold and heat at float.
Finally, a layer of light-absorbing paint or foil is applied to suppress stray light within the telescope, to lower measurement backgrounds.
This structure has been incorporated into paneling, which covers the Telescope Frame and Mirror Assembly. 

\paragraph*{Mirror Assembly}

\begin{wrapfigure}{r}{0.5\textwidth}
    \centering
    \includegraphics[width=0.48\textwidth]{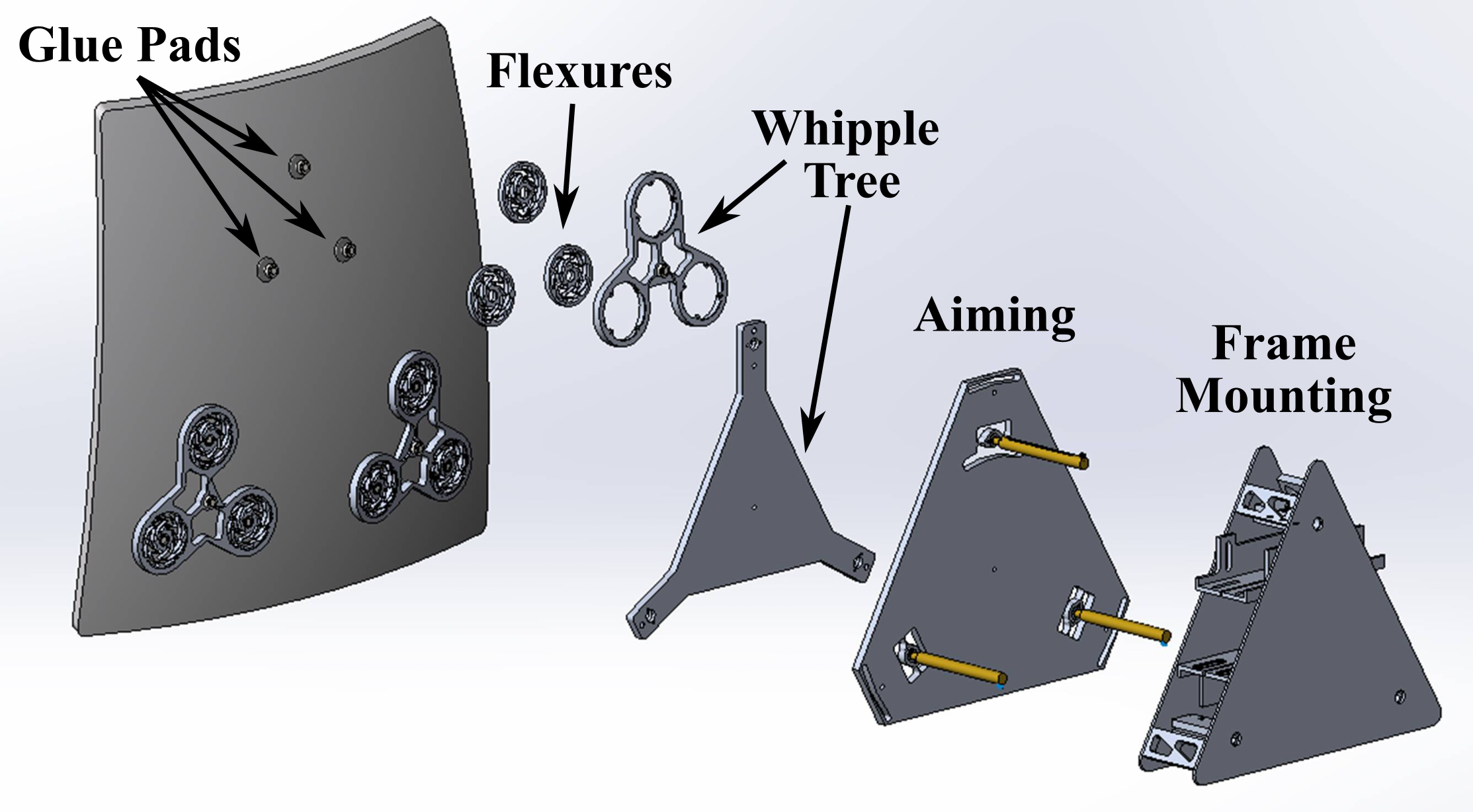}
    \caption{Mirror Cell}
    \label{fig:MirrorCell}
\end{wrapfigure}
The mirror assembly, shown in \autoref{fig:Mirror}, consists of 12 mirror cells and the mirror support cage.
A conceptual sketch of a mirror cell is shown in \autoref{fig:MirrorCell}.
Working left to right, first, there is the mirror segment.
Next are nine pads made of Kovar, which are glued to the mirror segment using Scotch weld D2216 Epoxy and 3901 primer.
Kovar alloy and the specific epoxy have the same or similar thermal properties as the glass in the mirrors, meaning the bond to the glass remains strong over the full range of temperatures faced by the payload. 
To attach the pads to the rest of the Aluminum assembly, flexures are used, which have been designed to absorb the force due to the thermal contraction difference between aluminum and the glass of the mirrors.
The flexures are placed into a Whipple tree assembly to distribute loads equally among the nine pads.
Targeted testing of the pad/glass bond shows it reliably holds more than 400\,lbs (181.4\,kg) at $-60^\circ$\,C, well below the lowest temperature expected for the mirror.
In FEA, the maximum per pad load seen under a $100^\circ$\,C $\Delta T$ and an 8\,G shock is 45.7\,lbs (20.7\,kg).
Next, there is the aiming plate, which allows for the mirrors to be steered to achieve the desired focus on the cameras.
Lastly, there are the frame mounting plates that secure the mirror cell to the mirror support frame.

\paragraph*{Camera Shelf}

The last mechanical component of the \ac{PBR} Telescope is the camera shelf.
The camera shelf serves as the interface point between the telescope frame and the \ac{FC} and \ac{CC}.
It will allow the two cameras to be independently moved $\pm 2.5$\,cm along the optical access while minimizing obscuration of the cameras or mirror.
Its design is still being finalized as the geometries and support structures of the \ac{FC} and \ac{CC} are still being designed.
\section{PBR Radio Instrument Mechanics}\label{sec:RI}

The radio instrument and assembly can be seen extending from the bottom of the telescope in \autoref{fig:Telescope}.
The main mechanical requirements on the radio assembly are: that it must fix the orientation of the radio antenna so that their normal vectors are parallel to the optical axis of the telescope within $1^\circ$, that the assembly survives an 8\,G shock with a safety factor of 1.5, that the radio is free from obscuration of any metal components within its $60^\circ \times 60^\circ$ \ac{FoV}, that the total assembly weighs no more than 150\,lbs (68\.kg), and that the whole assembly is free from conductive components save for nuts, bolts and dedicated radio electronics. 

The most challenging aspect of meeting all these requirements was identifying procurable non-conductive materials that could withstand these loads at $-70^\circ$\,C within the budget.
Ultimately, a combination of materials was necessary.
The mast assembly, which extends from the telescope to the radio antenna, is to be constructed from \ac{GFRV}, which has excellent strength properties and is readily available in the required lengths and geometries.
At joints, due to concerns with possible tear out, \ac{G10} gussets and bolt plates are employed.
The Radio Antennas are printed in copper on \ac{FR4} sheets, which have been sandwiched onto an Aramid honeycomb to provide stiffness. 
This combination yields very light, yet stiff, antenna assemblies with exceptional gain.
The frame that supports the antenna is a composite structure, consisting of \ac{FR4} blocking incorporated into the antenna panels themselves, paired with \ac{GFRV} bars to provide the required inter-antenna separation.
The amplification electronics are bolted directly onto the back of the antenna panels.
The full assembly is bolted directly to the telescope frame, providing a stiff and lightweight assembly that meets all requirements.
\section{Outlook}

The development and construction of \ac{PBR} is on track for an early 2027 flight.
The mechanical design of the \ac{PBR} payload is nearing completion, and the project is now moving to the fabrication and construction phase.
The majority of components will be ready by the end of 2025 and will then be sent to the Colorado School of Mines in Golden, Colorado, for integration during the Spring of 2026.
After integration, the telescope will be field-tested in Utah at the Telescope Array site ~\cite{Fukushima:2003ig} to profile the performance of the instrumentation in the Summer of 2026.
After field tests, in the Fall of 2026, \ac{PBR} will be sent to the \ac{CSBF} facility in Palestine, Texas, for hang testing and final flight approval.
From \ac{CSBF}, \ac{PBR} will be shipped to Wanaka, New Zealand, for flight on a super-pressure balloon in early 2027.
When flown, \ac{PBR} will make novel astroparticle observations from the stratosphere and will represent a significant step towards the proposed next-generation space-based \ac{UHECR} / \ac{VHENu} Observatory \ac{POEMMA}~\cite{POEMMA:2020ykm}.

\bibliographystyle{include/JHEPNoTitle.bst}
{\footnotesize
\bibliography{include/References.bib}{}
}

\vspace{-1ex}
\footnotesize
\section*{Acknowledgments}

The authors would like to acknowledge the support by NASA award 80NSSC22K1488 and 80NSSC24K1780, by the French space agency CNES and the Italian Space agency ASI.
The work is supported by OP JAC financed by ESIF and the MEYS CZ.02.01.01/00/22\_008/0004596. 
We gratefully acknowledge the collaboration and expert advice provided by the PUEO collaboration. 
We also acknowledge the invaluable contributions of the administrative and technical staffs at our home institutions.

\clearpage
    \newpage
{\Large\bf Full Authors list: The JEM-EUSO Collaboration}

\begin{sloppypar}
{\small \noindent
M.~Abdullahi$^{ep,er}$              
M.~Abrate$^{ek,el}$,                
J.H.~Adams Jr.$^{ld}$,              
D.~Allard$^{cb}$,                   
P.~Alldredge$^{ld}$,                
R.~Aloisio$^{ep,er}$,               
R.~Ammendola$^{ei}$,                
A.~Anastasio$^{ef}$,                
L.~Anchordoqui$^{le}$,              
V.~Andreoli$^{ek,el}$
% ...
}
\end{sloppypar}
\vspace*{.3cm}

{ \footnotesize
\noindent
%
% ...
$^{cb}$ Universit\'e de Paris, AstroParticule et Cosmologie (CNRS), Paris, France\\
% ...
$^{ef}$ Istituto Nazionale di Fisica Nucleare (INFN), Sezione di Napoli, Naples, Italy\\
% ...
$^{ei}$ Istituto Nazionale di Fisica Nucleare (INFN), Sezione di Roma Tor Vergata, Rome, Italy\\
% ...
$^{ek}$ Istituto Nazionale di Fisica Nucleare (INFN), Sezione di Torino, Turin, Italy\\
$^{el}$ Universit\`a di Torino, Dipartimento di Fisica, Turin, Italy\\
% ...
$^{ep}$ Gran Sasso Science Institute (GSSI), L’Aquila, Italy\\
$^{er}$ Istituto Nazionale di Fisica Nucleare (INFN), Laboratori Nazionali del Gran Sasso, Assergi, Italy\\
% ...
$^{ld}$ University of Alabama in Huntsville, Huntsville, AL, USA\\
$^{le}$ City University of New York (CUNY), Lehman College, Bronx, NY, USA\\
% ...
}

\end{document}